\documentclass[twocolumn,traditabstract]{aa} 

\usepackage{graphicx}
\usepackage{amsmath}
\usepackage{amssymb}
\usepackage{txfonts}
\usepackage{booktabs}
\usepackage{caption,natbib}
\usepackage[dvipsnames]{xcolor}
\usepackage[hidelinks]{hyperref}
\hypersetup{
    colorlinks=true,
    citecolor=blue,
    filecolor=black,
    linkcolor=blue,
    urlcolor=magenta,
    linktocpage=true,
    breaklinks = true
}

\newcommand{\bands}{\textit{griz}}
\newcommand{\lsne}{LSNe\,Ia}

\begin{document}

\title{OASIS: On-the-fly ASynchronous Image Stream classification}
\subtitle{Transformer-based models for the early detection of strongly lensed supernovae}

\author{Satadru Bag\inst{1,2}\thanks{satadru.bag@tum.de}, Raoul Ca{\~n}ameras\inst{3}, Sherry H.~Suyu\inst{1,2}, Stefan Schuldt\inst{4,5,6}, Stefan Taubenberger \inst{1,2}, Irham Taufik Andika\inst{7,1,2}, Alejandra Melo\inst{8,9}
          }

   \institute{Technical University of Munich, TUM School of Natural Sciences, Physics Department,  James-Franck-Stra{\ss}e 1, 85748 Garching, Germany
         \and 
           Max-Planck-Institut f{\"u}r Astrophysik, Karl-Schwarzschild Stra{\ss}e 1, 85748 Garching, Germany
           \and
          Aix Marseille Univ, CNRS, CNES, LAM, Marseille, France
           \and
          Finnish Centre for Astronomy with ESO (FINCA), University of Turku, FI-20014 Turku, Finland
          \and
          Department of Physics, P.O. Box 64, University of Helsinki, FI-00014
Helsinki, Finland
          \and
          INAF - IASF Milano, via A. Corti 12, I-20133 Milano, Italy
          \and
          University Observatory, LMU Faculty of Physics, Scheinerstrasse 1, 81679 Munich, Germany
          \and
          Departamento de F\'isica, Universidad de Santiago de Chile, Av.\,Victor Jara 3493, Santiago, Chile
          \and
          Center for Interdisciplinary Research in Astrophysics Space Science (CIRAS), Universidad de Santiago, Chile
}

\date{Received --; accepted --}

  \abstract
{
Time-domain surveys often do not deliver neatly sampled image sequences. Observations can
arrive at irregular times and in different photometric bands. In many
applications, a decision must be made after each new observation, particularly
when transient or variable sources must be identified early enough to trigger
follow-up. Strongly lensed supernovae (LSNe), for which follow-up can be particularly
time-critical, provide a useful example.
We formulate this setting as \emph{On-the-fly ASynchronous Image Stream} (\texttt{OASIS}) classification. We explore transformer architectures that organize the information in such streams in different ways. Our aim is not to identify a universally optimal architecture, but to provide reusable designs that can be adapted to different time-domain datasets and science cases. We demonstrate these architectures on the lensed-transient classification problem studied by Bag et al.\ (2026a,b), which aims at distinguishing LSNe Ia from unlensed astrophysical transients. We evaluate the transformer models against the recurrent baseline adopted by Bag et al. (2026b). The baseline processes images with a convolutional long short-term memory (\texttt{ConvLSTM}) branch and observation times with a separate \texttt{LSTM} branch, combining their outputs for classification. To accommodate asynchronous multi-band observations, it arranges the input on a fixed band-by-observation grid, filling the image entries for unobserved bands with zeros. We explore two broad ways of organizing interactions among the spatio-temporal tokens in these streams: \emph{factorized} attention, which treats spatial and temporal relationships separately, and \emph{joint} attention, which models them together. The
architectures are trained with a common setup so that their differences
primarily reflect how they organize the observations.
On the lensed-transient search dataset, the factorized architectures achieve
losses of $0.095$--$0.100$ on average across independent initializations,
compared with $0.128$ for the recurrent baseline and $0.107$ or higher for the
joint architectures. A selective causal readout further improves the
factorized representation, reaching a best loss of $0.089$, but provides no
corresponding benefit for joint attention. Adding a dedicated mechanism to combine information across photometric bands likewise does not improve performance.
Factorized attention provides the best performance in this demonstration, indicating that separating spatial and temporal interactions is a useful match to this particular dataset and science case. We do not regard this as a universal architectural choice: the appropriate design should be optimized for the science goal and data at hand. We therefore release the architecture families as reusable foundations for the wider time-domain community, and beyond astronomy for applications in which spatial features evolve over time and predictions must be made causally as observations arrive.
}
   {}
   {}
   {}
   {}

\keywords{methods: data analysis --
          methods: statistical --
          techniques: image processing --
          gravitational lensing: strong --
          supernovae: general}

\titlerunning{\texttt{OASIS}: Causal Classification of Asynchronous Image Streams}
\authorrunning{Bag et al.}

\maketitle
\section{Introduction}
\label{sec:intro}

Time-domain astronomy often relies on observations that arrive as a stream rather than as a complete, uniformly sampled time series. Images are obtained at irregular times and in different photometric bands, with the relative fluxes between bands providing important color information. A general prediction framework must therefore combine the observations available at each stage while accounting for their irregular temporal ordering and differing photometric bands.

We refer to this setting as \emph{On-the-fly ASynchronous Image Stream} (\texttt{OASIS}) classification. The observations are images, so spatial structure carries information; this
structure evolves with time; different bands need not be observed
simultaneously; and a decision may be required after every new observation.
These properties occur naturally in time-domain surveys, and their importance
is likely to grow as surveys provide larger and more diverse streams of
observations.

A particularly important example is the search for rare transient events, for which early identification can determine whether useful follow-up observations are possible. Strongly lensed supernovae (LSNe) are an important class of such transients. Strong gravitational lensing occurs when the gravitational potential of a massive galaxy or galaxy cluster bends light from a background source, producing multiple magnified images and, for extended sources, characteristic arcs and distortions \citep{1996astro.ph..6001N}. Signals from LSNe are intrinsically spatio-temporal: multiple images of the same supernova (SN) appear at different positions, with each image following the SN light curve at a different time delay and magnification determined by the lensing configuration \citep{Treu2010, Suyu_2024, LSNe_rev2}. Thus, the spatial arrangement of the images is linked to their temporal brightness evolution, providing a characteristic signature of lensing. If the lensed host galaxy is detectable in the neighborhood, it provides additional spatial information, while observations in different photometric bands constrain the color information. The evidence for a lensed event is therefore distributed across the spatial structure of the images and their evolution over time, and may not be contained in any single observation. Early identification requires a method that can capture and combine these signals as the observations accumulate.

LSNe are a scientifically important case in which early detection is particularly valuable. The time delays between their multiple images provide an independent measurement of the Hubble constant through time-delay cosmography (TDC; \citealt{Refsdal1964_1, Refsdal1964_2, Oguri_2007, 2016A&ARv..24...11T, Wong:2019kwg, 2020A&A...643A.165B, Birrer:2020jyr, Treu:2022aqp, Suyu_2024, TDCOSMO2025}), thereby helping to address the current Hubble-tension debate \citep{Planck:2018vyg, Riess2022,valentino2021,verde2024}. Accurate time-delay measurements benefit from well-sampled light-curve peaks, which early detection can facilitate through targeted follow-up observations. Early identification of the first image also allows the lens to be modeled and subsequent images to be predicted, enabling targeted observations of the later images and providing an opportunity to study the SN progenitor \citep{suyu20}. LSNe remain rare, with only a handful discovered so far. However, the next generation of wide-field surveys, such as the recently started Vera C. Rubin Observatory's Legacy Survey of Space and Time (LSST) \citep{lsst1,ivezic19} and the Roman Space Telescope in the future \citep{roman2015, Pierel2021}, is expected to increase their discovery rate substantially. LSST alone is predicted to discover hundreds of LSNe over its ten-year survey, along with thousands of lensed quasars \citep{om10,Nikki_2023,Bag:2024kbk}, which are also important for TDC. The ability to identify LSNe early in their evolution is therefore increasingly important, and the same \texttt{OASIS} framework may also be applicable to lensed quasars, including systems that are difficult to identify in static images.

Several strategies have been developed to identify LSNe, e.g., using cues from multiple images \citep{om10}, magnification \citep{goobar17,goldstein19,sagues24}, colors \citep{Quimby2014}, light curves \citep{Bag:2020pbg,misha2021,misha2022,Bag:2024kbk,Park:2026hmc}, or associations with known strong lenses \citep{shu18,craig24}. However, these approaches generally exploit only selected aspects of the available information and/or target specific subsets of lensed transients. A more general approach should combine the full spatial, temporal, and multi-band information in the image stream, enabling broader sample selection and more complete use of the data. Given the enormous data volumes of current and forthcoming wide-field surveys, automated machine-learning (ML) methods are well suited to such a task. Our earlier work developed such a deep-learning method for the early detection of LSNe \citep{bag2026a}. It processes the image sequence through a \texttt{ConvLSTM} branch \citep{shi2015}, which carries spatial information forward as the images are processed in time, while a second \texttt{LSTM} \citep{hochreiter1997} branch processes the corresponding observation times.\footnote{Previous studies have combined CNN-based image processing with recurrent networks applied to extracted light curves \citep{Morgan2022,Morgan2023}. \texttt{ConvLSTM}-based approaches have also been explored using simulated single-band data \citep{Ramanah2022}; our implementation instead processes multi-band image sequences and is trained on simulations based on real survey observations.} The model produces an updated prediction after every observation, naturally enabling causal, on-the-fly classification, often termed ``online classification'' in ML terminology. We use this model as a strong recurrent baseline rather than as a straw man.

Its main difficulty is the asynchronous multi-band structure of the data. The
\texttt{ConvLSTM} expects the bands to be represented as the channels of an image, so
the observations must first be arranged on a regular band-by-observation grid.
With one band observed at each step, the remaining band entries are filled with
zeros. For four bands, three of the four entries associated with each
observation are therefore empty. The recurrent architecture also processes
the stream sequentially, carrying information forward through its recurrent
state.

Transformer architectures \citep{vaswani2017} offer a fundamentally different approach. Their ``attention'' mechanism \citep{Bahdanau14,Luong2015,Parikh16} allows information to be selectively exchanged between different parts of the input, represented as \emph{tokens}. In our setting, this allows information from different observations and spatial regions to be related directly, without requiring the observations to be arranged on a fixed band-by-time grid. This provides a flexible framework for representing and processing the asynchronous observation stream, with several potential advantages for our problem. In particular:

\begin{itemize}
\item \textbf{Flexible representation.}
The observations can be treated as the data that actually arrived, with their
band and arrival time explicitly encoded. Different attention patterns can
then be used to decide how spatial and temporal information should interact.
The architecture is therefore not tied to one particular way of representing
an asynchronous stream, and further attention-based designs can be developed
for different surveys and science cases.

\item \textbf{Efficient parallel computation.}
Unlike recurrent processing, attention can evaluate many interactions in
parallel. This makes transformer architectures well suited to modern
high-performance accelerators and allows computational resources to be spent
on larger models or more elaborate patterns of information exchange when the
problem warrants them.

\item \textbf{An actively developing architecture family.}
Transformers have evolved rapidly beyond their original applications in
language, with attention-based methods now being developed for images, video,
irregular data, and other forms of structured information. This continuing
development provides a natural opportunity to improve or adapt the
architectures for time-domain streams, rather than treating the network
design as a fixed solution.
\end{itemize}

The flexibility of attention, however, also creates a new question: how should attention be organized for an asynchronous sequence of observations? One possibility is to separate spatial and temporal interactions, so that the model first relates different spatial regions within each observed image and then relates their representations across observations. Another is to allow spatial and temporal information to
interact directly. We refer to these as \emph{factorized} and \emph{joint}
attention, respectively. Within these two broad classes, different choices
can also be made about the spatial receptive field and about how the
information accumulated so far is condensed into the prediction.

Our goal is not to benchmark these architectures or identify a universally best transformer. The appropriate choice depends on the survey cadence, number of bands, image properties, data volume, and scientific objective. Instead, we use the LSNe-search problem as a concrete demonstration of \texttt{OASIS}, exploring a small set of architectural families that represent different ways of processing asynchronous image streams. All architectures considered here are made publicly available in the accompanying repository.

Although developed with astronomical time-domain surveys in mind, \texttt{OASIS} addresses a broader problem: making predictions from spatially structured observations that arrive asynchronously across multiple channels. The same setting arises beyond astronomy whenever spatial features evolve over time and decisions must be made on-the-fly. The architectures explored here may therefore also serve as starting points for asynchronous multi-channel image-stream applications in other fields.

The remainder of this paper is structured to keep the scientific argument accessible in the main text without requiring a background in ML, while the implementation details and technical ML aspects are left to the appendices. Sec.~\ref{sec:problem} formulates the \texttt{OASIS} problem and identifies the four properties that distinguish it from conventional time-series classification. Sec.~\ref{sec:data} introduces the LSNe demonstration dataset. Sec.~\ref{sec:models} presents the convolutional-recurrent baseline and attention-based architectures, and Sec.~\ref{sec:results} examines how different ways of organizing information exchange among tokens across space and time affect performance, as well as the effects of readout and cross-band fusion. Sec.~\ref{sec:discussion} draws out the broader architectural lessons, and Sec.~\ref{sec:conclusions} summarizes the main conclusions. Complete model specifications and the training protocol are given in Appendix~\ref{app:arch}.

For the demonstration dataset, we use time series of $59\times59$-pixel image cutouts, corresponding to a field of view of approximately $10\arcsec\times10\arcsec$, and assume a flat $\Lambda$CDM cosmology with $\Omega_{\rm M}=0.308$ \citep{planck16} and $H_0=72\,{\rm km\,s^{-1}\,Mpc^{-1}}$ \citep{Bonvin2017}.


\section{The problem}
\label{sec:problem}

\subsection{Definition}

We consider a sequence of observations in $N_{\rm B}$ photometric bands (for the LSNe
demonstration, $N_{\rm B}=4$, the \bands\ filters). The observations form a
time-ordered stream,
\begin{equation}
S = \bigl[(x_1,b_1,t_1),(x_2,b_2,t_2),\dots\bigr],
\label{eq:stream}
\end{equation}
where the $k$-th observation consists of an image $x_k$ obtained in band
$b_k\in \lbrace 1,\dots,N_{\rm B} \rbrace$ at time $t_k$, with
$t_1\le t_2\le\dots$. The observations are asynchronous: one band is observed at each step in general\footnote{When multiple bands are observed during the same night, the observations get closely spaced timestamps.}, the intervals between observations are irregular, and no interpolation is used to construct observations that were not obtained.

\begin{figure*}[t]
\centering
\includegraphics[width=\textwidth]{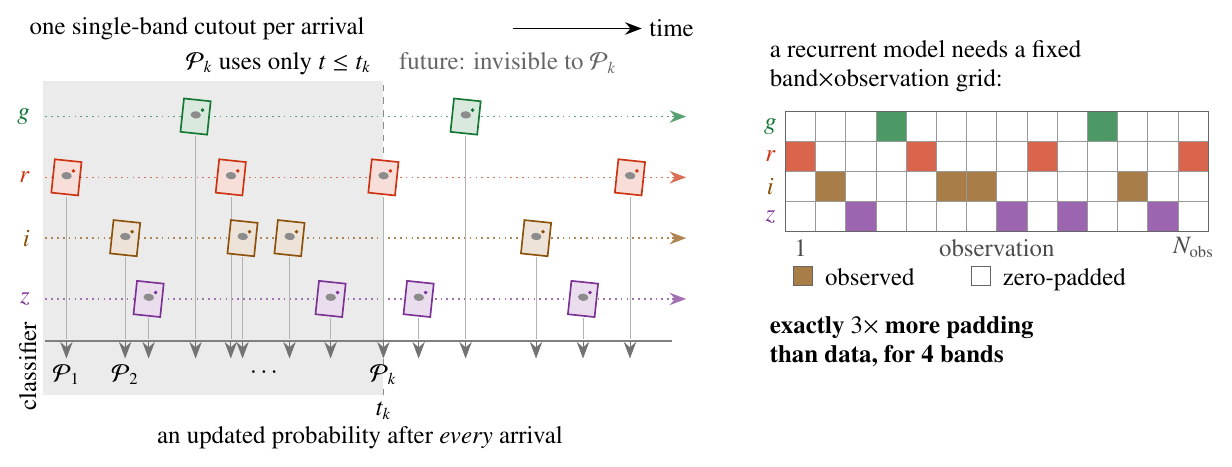}
\caption{The \texttt{OASIS} problem. Single-band cutouts arrive one at a time at
irregular intervals, and an updated probability $\mathcal{P}_k$ is emitted
after each arrival using only the observations available at that time
(shaded). Here $N_{\rm obs}$ denotes the total number of observations in the
stream. \emph{Right:} the same stream placed on a fixed
band$\times$observation grid, as required by the recurrent model used in
our baseline. With one band arriving per step, this representation contains
three times as many empty entries as observed images.}
\label{fig:problem}
\end{figure*}

For our LSNe demonstration, the task is binary classification: after each observation, using only the observations available up to that point, the classifier produces an updated probability
\begin{equation}
\mathcal{P}_k =
f\bigl((x_1,b_1,t_1),\dots,(x_k,\,b_k,\,t_k)\bigr) \in [0,1]
\label{eq:online}
\end{equation}
that the sequence contains an LSN system (positive class). Predictions are thus made causally and on-the-fly. Fig.~\ref{fig:problem} illustrates the \texttt{OASIS} problem, showing the resulting multi-channel observation stream and the causal constraint.

\subsection{Four difficulties, and why they are general}
\label{sec:Ds}

Four properties make \texttt{OASIS} different from a conventional time-series classification problem. We summarize these as four design requirements, denoted D1--D4, that a model for this problem should satisfy:

\emph{(D1) The primary data are \textbf{images} with spatial correlation (features), not numbers.}
The relevant information is both \emph{spatial}, such as the morphology and
relative positions of sources, and \emph{temporal}, such as how these features
change from one observation to the next. The model therefore needs to capture
spatio-temporal structure rather than only temporal evolution of a set of scalar measurements.

\emph{(D2) The bands arrive \textbf{asynchronously}.}
Only one band is observed at each step, so no naturally co-registered multi-band image is available at any point in the stream. Consequently, treating the bands as channels of a single image requires a band-by-observation grid with padded entries, most of which contain no observation, as illustrated in the right panel of Fig.~\ref{fig:problem}. This motivates treating the data as an asynchronous stream rather than using such a padded representation.

\emph{(D3) Information between bands must still be combined at
\textbf{fine spatial scales}.}
Although the bands arrive separately, their relationships, e.g., color information encoded in flux ratios between bands, can carry important information at the level of individual pixels or small spatial regions. The model therefore needs to
combine information from different bands while retaining spatial detail,
rather than processing each band independently and combining only
whole-image summaries.

\emph{(D4) Predictions are \textbf{made on-the-fly and causally}.}
A prediction is required after every observation, using only the observations
available at that point in the stream. The number of observations available
to the model can therefore change from one prediction to the next, and the
model must remain applicable as the stream grows. Information from later
observations cannot be used to improve an earlier prediction.

These properties are characteristic of many time-domain survey problems, including
but not limited to the search for lensed transients. More generally, whenever
spatial features evolve over time and predictions must be made causally as new
observations arrive, the problem has the same structure. The architectures
explored below are therefore intended as general starting points for such
applications, rather than solutions specific to the astrophysics of lensing.

\section{Demonstration dataset}
\label{sec:data}

We stress that the dataset used here is a \emph{demonstration} chosen to make
the \texttt{OASIS} setting concrete. It is taken from \citet{bag2026b} and used
without modification; the contribution of this paper is the formulation of
the problem and the exploration of architectures for addressing it.

\subsection{Science case}

The demonstration is formulated as a binary classification problem: identifying a stream of time-domain images containing a galaxy-scale lensed Type-Ia SN (LSN~Ia), whose light is multiply imaged by a foreground lens galaxy, while rejecting streams containing unlensed transients or other variable sources that can exhibit observational signatures resembling those of LSNe. Early identification, ideally before the first-arriving image reaches peak brightness, enables targeted follow-up and precise time-delay measurements, giving these systems significant cosmological and astrophysical value. 

\subsection{Data}

We use the \bands\ image time series from our earlier work, \citet{bag2026b}, generated with the \texttt{GLEE}-based simulation pipeline \citep{glee1, glee2, schuldt21}, without modification, and refer the reader there for the full construction. The main purpose of this
dataset is to construct LSST-like time-domain data for training LSNe search
models. Rather than simulating the observations entirely from scratch, we
ingest real observations to retain the complexities of real survey data
in the training set. As a preparation for LSST before its time-domain data
become available, we use observations from the Hyper Suprime-Cam (HSC) survey
\citep{aihara18, yasuda19} as a proxy for LSST. Although LSST observes in
six bands, $ugrizy$, we restrict our analysis to the four bands $griz$, for
which the HSC data products used here provide sufficiently complete and
consistent coverage. The $u$ band is nevertheless expected to have relatively sparse cadence in LSST.

The positive
class consists of simulated galaxy-scale \lsne\ systems: the multiple SN
images and the lensed arcs of the SN host galaxy are
injected into \bands\ cutouts of real luminous red galaxies (LRGs) observed by
HSC. The negative class
has several components: real variable sources from the HSC Transient Survey
\citep[HTS;][]{yasuda19,chao21}, along with unlensed SNe~Ia placed in three environments: the foreground
galaxy of a genuine lens system, LRGs, and spiral galaxies. For a detailed
description of the dataset, we refer the reader to
\citet{bag2026a,bag2026b}.

Relative to \citet{bag2026a}, the simulations incorporate several additional
sources of realism, most importantly epoch-to-epoch variation in
the point-spread function (PSF) with consistently corrected variance maps, together
with Poisson noise. The light curves include intrinsic stretch and color
variation. Each system contains $N_{\rm obs}=14$ observations for the controlled
comparison performed here, distributed asynchronously over the four bands,
one band per observation, following the cadence of HTS. This fixed number of
observations is a choice for this comparison, rather than a requirement of the
\texttt{OASIS} problem. The time stamp of the first detection defines $t=0$, and the
full sequence spans at most 120 days. The training set contains $128{,}000$
samples, half of them positive; the validation and test sets, which share the
same class composition, contain $16{,}000$ and ${\sim}88{,}600$ samples,
respectively, with ${\sim}8.2\%$ positives \citep[see Table~1 of
][]{bag2026b}.

Two features of the data are worth noting. First, the training set deliberately over-samples the hardest false positives, whereas the validation set has the same composition as the test set. Second, the inputs are normalized per sample relative to the first arriving image, preserving the flux ratios between bands \citep{bag2026a}. This normalization is part of the published dataset and is left unchanged here.


\section{Baseline and architectures}
\label{sec:models}

\subsection{The convolutional-recurrent baseline}
\label{sec:models:convlstm}

We use the convolutional-recurrent architecture introduced by
\citet{bag2026a} as the baseline. The model has two recurrent components with
different roles. A convolutional recurrent layer (\texttt{ConvLSTM}; \citealt{shi2015})
processes the images in temporal order, carrying forward spatial features as
the observations evolve. In parallel, a recurrent layer (\texttt{LSTM};
\citealt{hochreiter1997}) processes the corresponding observation times,
allowing the model to learn temporal scales that can differ between classes.
The outputs of these two components are combined to produce an updated
$\mathcal{P}_k$ after each observation. In its original formulation, the
available photometric bands are stacked as the channels of an image. This allows the convolutional layers to capture spatial structure (D1)\footnote{These design requirements (D1-D4) are defined in Sec.~\ref{sec:Ds}.}, while the two recurrent components capture the evolution of the image features and observation times (D4). Channel stacking also provides a natural way to combine information between bands at the pixel level (D3).

The main difficulty is the asynchronous arrival of the bands (D2). Channel
stacking presupposes a multi-band image at each step, whereas the stream
contains only one newly observed band. In \citet{bag2026a}, the missing bands
are therefore filled with zeros before the images are passed to the network.
For four bands and one real image per observation, three quarters of the
resulting band-by-observation representation contain no observed image. The
padding is not only computationally redundant: it forces the convolutional
network to operate on a representation dominated by empty entries, while the
pixel-level fusion between bands is performed through a mixture of observed
and zero-filled inputs.

A seemingly cleaner alternative would be to process each photometric band with a separate recurrent branch and fuse the resulting features whenever a new observation arrives. Each branch would then be updated only when its corresponding band is observed, avoiding the zero-filled entries in the channel-stacked representation. However, this separation delays the interaction between bands, potentially preventing the model from learning useful color information directly at the pixel level. More importantly, it introduces a difficulty during training. Backpropagation through time requires the samples in a batch to follow the same sequence of band arrivals, because the recurrent states and their gradients depend on the ordering of the observations. Although batches could be constructed with a common ordering, this would require separate handling of the many possible orderings. With $N_{\rm B}$ bands and $N_{\rm obs}$ observations, the number of possible orderings is $N_{\rm B}^{N_{\rm obs}}$. For example, four bands observed over 14 epochs give $4^{14}=268{,}435{,}456$ possible orderings. Thus, while separate recurrent branches provide an attractive way to represent asynchronous data, enforcing a consistent ordering during training becomes impractical for the observation streams considered here.

The baseline used here is a slightly deeper configuration of the architecture introduced by \citet{bag2026a}, adopted from \citet{bag2026b}.\footnote{The implementation is publicly available at \url{https://github.com/deltasata/ConvLSTM-LSNe-Detection}.} On the present demonstration dataset, the baseline achieves a mean
per-observation loss of $0.128$, with the best-seed model reaching $0.114$.

\subsection{An attention-based alternative}
\label{sec:models:why}

Attention \citep{Bahdanau14,Luong2015, Parikh16, vaswani2017} provides a natural alternative to the fixed band-by-observation representation. Instead of requiring observations to occupy predefined positions in a grid, the model can operate directly on the images that have actually arrived. Each image is divided into spatial \emph{patches}, with each patch represented by a token, a feature vector that also encodes its spatial position and the observation's time and band information. Each observation therefore contains $N_{\rm P}=81$ patch tokens, corresponding to the $9\times9$ spatial grid, and each system contains $N_{\rm obs}=14$ observations distributed across the four bands. The network then learns how information from these tokens should be combined across observations. A causal attention scheme ensures that, when producing $\mathcal{P}_k$, only observations up to $k$ can contribute. This removes the need for zero-padding the unobserved bands and allows information from different bands to be combined without requiring them to arrive simultaneously.

The main architectural question is then how attention should exchange information among tokens across space and time. We consider two broad classes. In
\emph{factorized} attention (Class~A), spatial and temporal relationships are
handled separately. The regions within an observation first exchange spatial
information, after which each spatial region can exchange information with
the corresponding region in earlier observations; these two operations are
alternated through the network. In \emph{joint} attention (Class~B), spatial
and temporal relationships are considered together, allowing regions at
different positions and different observation times to interact directly.

A second choice is how the information accumulated by the network is turned
into the classification probability. We call this the \emph{readout}. A
simple readout gives equal weight to the spatial regions of the current
observation. A selective readout instead learns which of the causally
available observations and spatial regions are most useful for the current
prediction. We also consider a richer version in which several learned
queries gather information in parallel.

From these choices we construct seven transformer-based architectures,
divided into two classes and summarized in Table~\ref{tab:headline}.
Class~A contains \textbf{A1}, with factorized attention and mean-pool readout;
\textbf{A2}, with the same factorized backbone and a single learned query for
the readout; and \textbf{A3}, which uses sixteen learned queries that are
refined together before their outputs are combined. Class~B contains
\textbf{B1}, with unrestricted joint attention; \textbf{B2w} and \textbf{B2d},
which restrict joint attention to a $3\times3$ spatial neighborhood while
retaining access to all previous observations, using wider and deeper
networks, respectively; and \textbf{B3}, which first applies local attention
on the fine spatial grid and then uses a pooled coarse grid to recover
larger-scale context. Table~\ref{tab:headline} highlights the main properties
of the seven architectures and the choices that distinguish them. The
attention patterns and complete architectural specifications are given in
Appendix~\ref{app:arch}.

\begin{table*}[t]
  \caption{The architectures compared as solutions to the \texttt{OASIS} problem described in
Sec.~\ref{sec:problem}, with their per-observation loss.}
  \label{tab:headline}
  \centering
  \renewcommand{\arraystretch}{1.25}
  \begin{tabular}{@{}l p{4.7cm} p{3.55cm} c c c c@{}}
    \toprule
    & & & & & \multicolumn{2}{c}{Test loss (lower is better)}\\
    \cmidrule(lr){6-7}
    ID & How space and time are attended & How $\mathcal{P}_k$ is read out
       & \shortstack{width, depth\\ $(d_{\rm model},N_{\rm layer})$}
       & Par.\ [$10^6$] & best & mean $\pm$ sd \\
    \midrule
    -- & \emph{Baseline} \citep{bag2026a, bag2026b}: \texttt{ConvLSTM} processes the image sequence;
\texttt{LSTM} processes the corresponding observation times
      & combined recurrent representations at every step
      & -- & $0.6$ & $0.114$ & $0.1277 \pm 0.0073$ \\
    \midrule
    A1 & \emph{Factorized}: the patches of one observation attend to one
      another, then each patch looks back along its own history,
      alternately
      & \emph{mean pooling}: the current observation's tokens are averaged
        with equal weights
      & $256,\,4$ & $4.0$ & $0.0947$ & $0.0999 \pm 0.0040$ \\
    A2 & as A1
      & \emph{one learned query} per observation gathers all causally available tokens with learned weights
      & $256,\,4$ & $4.3$ & $\mathbf{0.0890}$ & $0.0952 \pm 0.0030$ \\
    A3 & as A1
      & \emph{latent array}: $16$ learned queries per observation, refined
        together over three rounds, then averaged
      & $256,\,4$ & $6.0$ & $0.0915$ & $\mathbf{0.0948 \pm 0.0031}$ \\
    \midrule
    B1 & \emph{Joint, global}: one causal attention over all
      $N_{\rm obs} N_{\rm P} = 1134$ space--time tokens; every visible token can attend to every other visible token
      & mean pooling
      & $256,\,4$ & $2.5$ & $0.1373$ & $0.1415 \pm 0.0044$ \\
    B2w & \emph{Joint, windowed}: as B1, but each token sees only its
      $3\times3$ patch neighborhood, now and in every past
      observation
      & mean pooling
      & $384,\,4$ & $4.5$ & $0.1043$ & $0.1084 \pm 0.0029$ \\
    B2d & as B2w, a deeper and narrower shape
      & mean pooling
      & $256,\,6$ & $3.3$ & $0.1011$ & $0.1072 \pm 0.0045$ \\
    B3 & \emph{Joint, pyramid}: windowed attention on the fine $9\times9$
      patch grid, then average-pooled to a coarse grid, where attention is unrestricted in space while remaining causal in time
      & \emph{time-conditioned readout}: the current observation's time embedding is used as the query
      & $256,\,3{+}3$ & $3.5$ & $0.1021$ & $0.1113 \pm 0.0048$ \\
    \bottomrule
  \end{tabular}
  \tablefoot{Each architecture was retrained from more than ten independent
random initializations; `best' is the lowest loss among these runs and `mean $\pm$ sd' gives their mean and standard deviation. `Width' ($d_{\rm model}$) is the length of each token's feature
    vector; `depth' ($N_{\rm layer}$) is the number of stacked layers, counting the alternating
spatial$+$temporal pairs for Class~A, the joint-attention layers for B1,
B2w and B2d, and the fine$+$coarse layers for B3.}
\end{table*}

\subsection{A controlled comparison}
\label{sec:models:training}

We selected the seven architectures from a broader exploration of the two
classes and evaluated each over more than ten independent random
initializations. This reduces the dependence of the comparison on any single
training run and allows us to assess the behavior of an architecture rather
than an individual model realization.

We kept the data, objective, optimizer, batch and patch sizes, regularization,
and time and band encoders fixed across the architectures. The objective is
the standard binary classification loss applied to the prediction at every
observation, so that early as well as late predictions contribute to training.
The only architectural difference in the controlled comparison is therefore
the network design; the complete training setup is given in
Appendix~\ref{app:arch:training}.

\section{Results}
\label{sec:results}

We evaluate the models using the per-observation binary cross-entropy loss on the fixed training, validation, and test sets, with the loss averaged over all observations in the complete time series. The training loss is used for parameter optimization (i.e. training), the validation loss for adaptive training and model selection, and the test loss only for the final evaluation. The training protocol is described in Appendix~\ref{app:arch:training}. The loss provides an overall measure of predictive performance over the full sequence, but does not characterize performance at different classification thresholds or directly quantify early detection capability. These aspects are examined separately in Sec.~\ref{sec:results:detection}.

We refer to the test loss simply as the ``loss'' below. The results apply specifically to the LSNe demonstration and the dataset of \citet{bag2026b}; they should therefore not be interpreted as a universal ranking of architectures. Relative performance may change with the science goal, survey cadence, number of bands, or other properties of the data. Our purpose here is instead to examine which architectural choices appear useful for the \texttt{OASIS} setting on this concrete example.

\subsection{Factorized attention provides a better match to the problem}
\label{sec:results:factorized}

Table~\ref{tab:headline} summarizes the loss of the seven
architectures. We use the table as a map of the architectural choices
explored in Sec.~\ref{sec:models}, rather than as a leaderboard. On this
demonstration, the most pronounced difference is between the two ways of
organizing attention to exchange information across space and time.

The three factorized architectures have mean losses between
$0.095$ and $0.100$, compared with $0.107$--$0.142$ for the joint
architectures and $0.128$ for the recurrent baseline. The best individual run
among the factorized architectures reaches $0.089$, but the more relevant
result is their consistently lower mean loss across the independent runs.

The joint architectures show a clear dependence on the spatial extent of
their attention. Global joint attention (B1) performs substantially worse
than the recurrent baseline, whereas restricting the spatial interaction to a
$3\times3$ neighborhood (B2w and B2d) brings the performance much closer to
the factorized models. Increasing the capacity through width or depth gives
very similar results, and adding a second spatial scale in B3 does not
provide a further improvement. Thus, within the joint class, restricting how
spatial information is combined appears more important than increasing model
capacity or adding a more elaborate spatial hierarchy.

Taken together, these results suggest that, for this particular dataset, separating spatial and temporal interactions is a more suitable organization of information than allowing them to interact everywhere at once. The following subsections examine two further architectural questions
that emerge from this comparison: how the accumulated information should be
read out, and whether cross-band information needs an explicit fusion
mechanism.

\subsection{The readout matters, but its effect depends on the backbone}

The readout determines how the information accumulated by the network is condensed into the prediction. On the factorized backbone, replacing the simple mean pooling of A1 with the learned selective readout of A2 improves the best loss from $0.0947$ to $0.0890$. This improvement does not require a large model: using the same selective readout with a width of $96$ yields a loss of $0.0935$ in one exploratory run, already below that of the width-$256$ mean-pool model.

Making the readout more elaborate provides little further benefit. A3, which
uses sixteen learned queries rather than a single query, gives a mean
loss of $0.0948$, comparable to the $0.0952$ of A2. Thus, once the
readout can selectively gather information from the causally available
observations, increasing the number of learned summaries does not appear to
provide a clear advantage on this dataset.

The same idea behaves differently on the joint representations. Across the global and windowed joint backbones we tested, the selective readout showed no consistent benefit, with changes generally comparable to the seed-to-seed variation of the mean-pool models. The deeper global variant was a clear exception, with the selective readout substantially worsening the loss. By contrast, on the factorized backbone, the improvement exceeded the seed-to-seed variation and was reproduced across runs. These results suggest that the benefit of selective readout depends on the representation produced by the preceding attention layers: it consistently improves the factorized model but offers no comparable benefit for the joint models.

\subsection{Explicit cross-band fusion does not provide an advantage}

The asynchronous arrival of the bands raises a natural question: should the
network be given an explicit mechanism to combine information between bands?
We explored several such modifications in addition to the seven architectures
in Table~\ref{tab:headline}, including attention between the most recent
images in different bands, per-band feature rescaling, and channel stacking
in the convolutional stem. None improved on the plain factorized
architecture. A further experiment that forced the temporal information
through a small set of persistent summary tokens performed worse still.

These experiments suggest that explicit color-fusion mechanisms are not
necessary for this problem. The factorized architecture can learn the
relationships between bands through its ordinary spatial and temporal
attention, without requiring a separate band-fusion operation. In this sense,
the organization of the attention appears to matter more than adding a
purpose-built mechanism for combining the photometric bands.

One additional experiment provides a useful nuance. Replacing the learned
band identifiers with the physical central wavelengths of the HSC \bands\ filters
($473$, $623$, $777$, $892$\,nm; \citealt{Kawanomoto2018}) improves the joint
architectures by $\mathcal{O}(10^{-3})$ in loss, while leaving the factorized
architectures essentially unchanged.
The learned band encoding already tells the network which filter produced each
observation, but does not explicitly provide the physical ordering of the
filters. Supplying the wavelengths therefore gives the joint architectures
additional information about the relationships between bands. The fact that
this has little effect on the factorized architectures suggests that they can
already recover the useful cross-band structure from the observations. Even
with the wavelength information, however, the best joint model does not reach
the factorized performance.

\subsection{Detection performance}
\label{sec:results:detection}

\begin{figure*}[t]
  \centering
  \includegraphics[width=\textwidth]{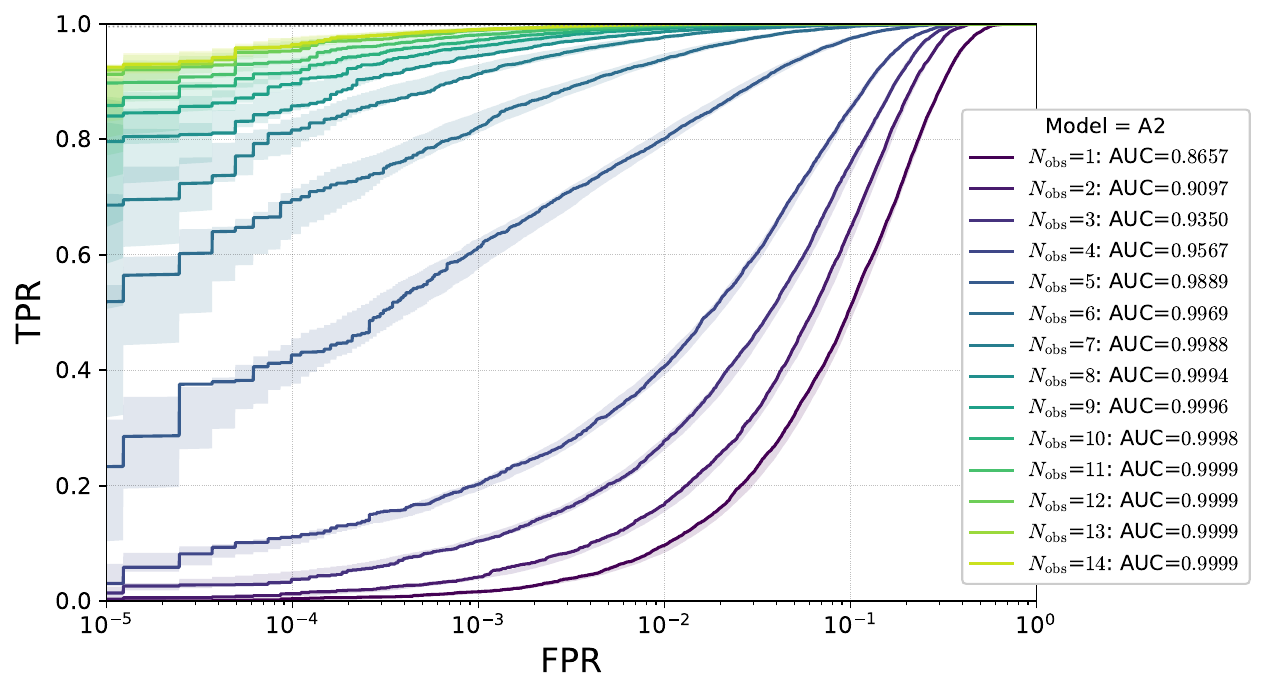}
  \caption{Test-set ROC curves for the A2 model, showing TPR as a function of FPR as the threshold on the predicted scores varies. Each curve corresponds to a different observation number, $N_{\rm obs}$; the solid curves show the best-seed A2 model and the shaded regions show the mean $\pm$ standard deviation across the independent A2 runs. The corresponding AUC values are given in the legend.}
  \label{fig:roc}
\end{figure*}

\begin{figure}[t]
  \centering
  \includegraphics[width=\columnwidth]{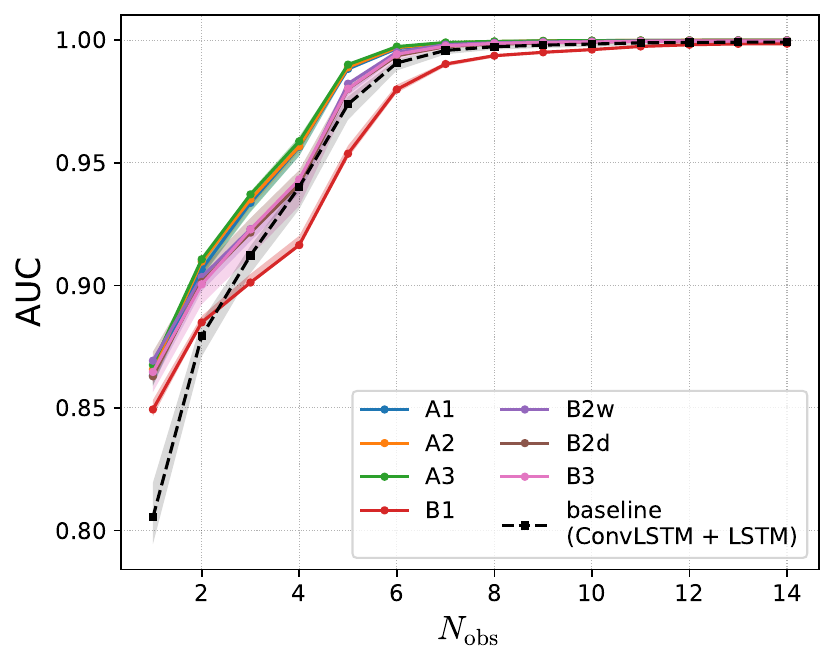}
  \caption{Test-set AUC as a function of the number of observations for the
  seven architectures considered in Table~\ref{tab:headline}. The dashed
  black curve shows the recurrent baseline (\texttt{ConvLSTM} + \texttt{LSTM}).}
  \label{fig:auc}
\end{figure}

\begin{figure*}[tbh]
  \centering
  \includegraphics[width=\textwidth]{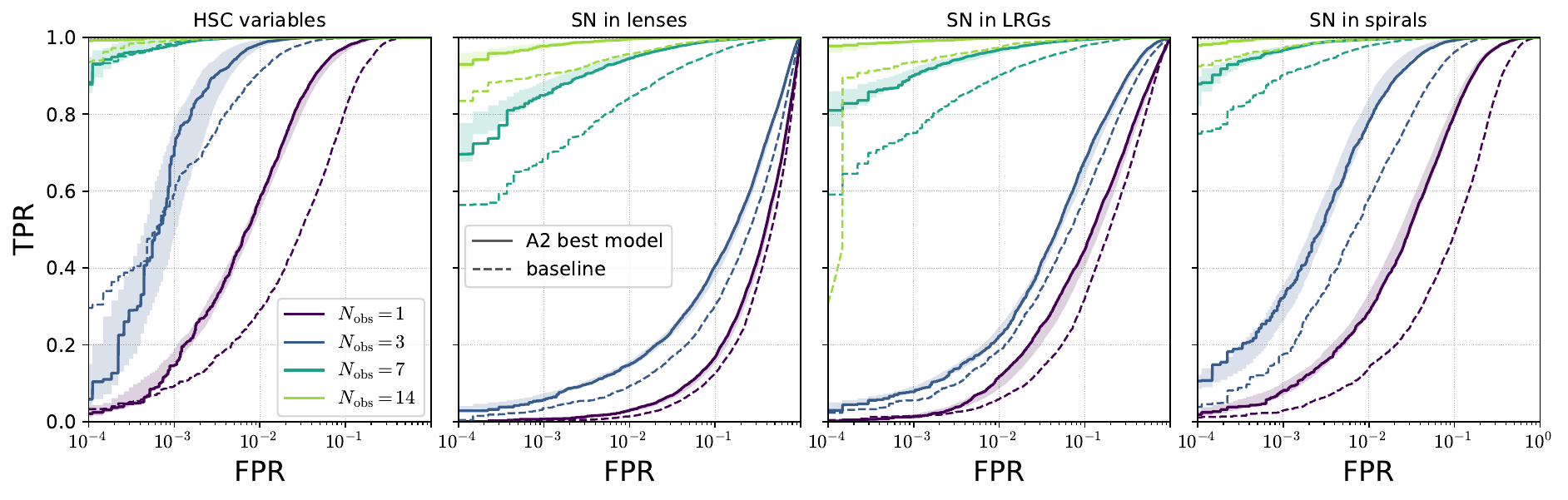}
  \caption{ROC curves for the A2 model at $N_{\rm obs}=1$, 3, 7, and 14 observations, shown separately for the four negative-source populations in the LSNe demonstration. Solid curves show the best-seed A2 and dashed curves the convolutional-recurrent baseline; the shaded regions indicate the mean $\pm$ standard deviation across the independent A2 runs. As observations accumulate, the classifier increasingly separates LSNe from all contaminant populations, although SNe in lens galaxies and LRGs remain the most challenging false positives.}
  \label{fig:roc_subclass}
\end{figure*}

\begin{figure}[t]
  \centering
  \includegraphics[width=\columnwidth]{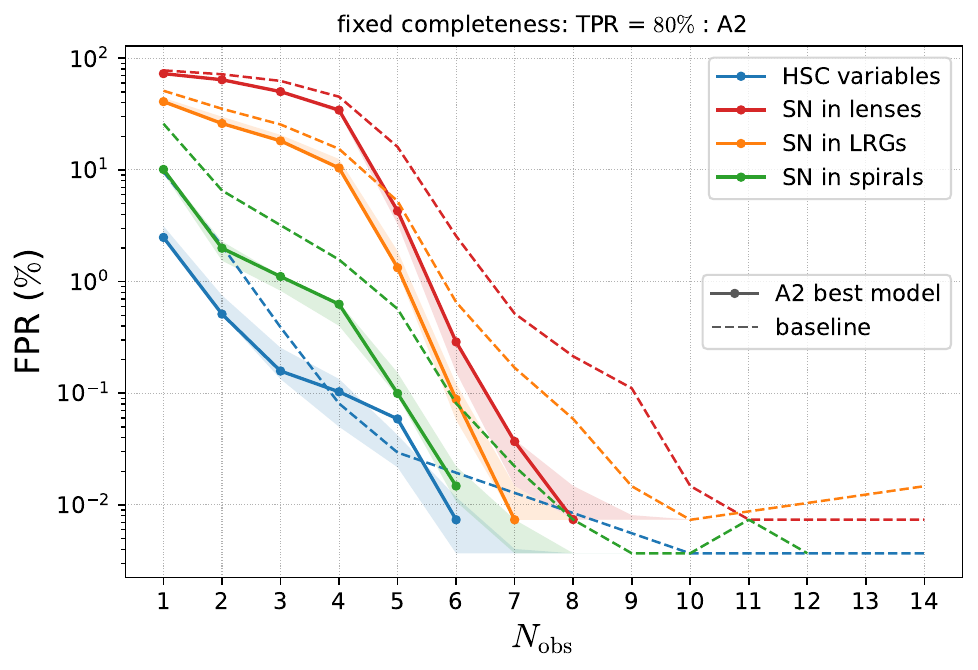}
  \caption{Test-set FPR at fixed TPR of $80\%$ as a function of the number of
  observations, shown separately for the different negative-source
  populations. Solid curves show the best-seed A2 model, dashed curves show
  the recurrent baseline, and shaded regions show the mean $\pm$ standard
  deviation across the independent A2 runs. The solid curves terminate when
  no false positives are observed in the corresponding population, indicating
  an empirical zero-FPR floor.}
  \label{fig:class_fpr}
\end{figure}

We next examine how the differences seen in the loss translate into detection performance on the held-out test set, including how performance changes as observations accumulate. We focus on A2, the best-seed factorized model selected from the validation runs, and compare it with the convolutional-recurrent baseline. 

After each observation $k$, the model produces a predicted score $\mathcal{P}_k$ according to Eq.~\eqref{eq:online}, based only on the observations available up to that point. Because the models are causal, this prediction is identical to that obtained from a stream truncated after the $k$-th observation. Thus, $N_{\rm obs}=k$ in Figs.~\ref{fig:roc}--\ref{fig:class_fpr} corresponds to the number of observations available at the time of prediction, and the same trained model can be applied to streams of different lengths. Fig.~\ref{fig:roc} shows the Receiver Operating Characteristic (ROC) curves of A2 as observations accumulate. The ROC curve shows the True Positive Rate (TPR) against the False Positive Rate (FPR) as the threshold applied to the predicted scores is varied.

The solid curves show the best-seed A2 model, while the shaded regions indicate
the mean $\pm$ standard deviation across the independent A2 runs. The area under the ROC curve (AUC) for each $N_{\rm obs}$ is given
in the legend. Even after a single
observation, A2 provides useful discrimination, with an AUC of 0.866, already
substantially better than the recurrent baseline. More importantly for early
detection, the classifier rapidly reaches high completeness at low
FPR: at $\mathrm{FPR}=10^{-4}$, the TPR exceeds
$\sim70\%$ by the sixth observation and $\sim80\%$ by the seventh.

Fig.~\ref{fig:auc} extends this comparison to all seven architectures
against the recurrent baseline. The AUC increases rapidly with the number of
observations for all models, reflecting the progressive accumulation of
information in the stream. The factorized architectures consistently provide
the strongest early-time discrimination, while the differences between models
become smaller as more observations accumulate.

This also illustrates why the number of observations is an important part of
the detection problem: a single final performance number does not describe how
useful a classifier is for triggering follow-up at different stages of a
transient.

AUC, however, averages over all classification thresholds and does not directly
show the FPR arising from different negative components. Fig.~\ref{fig:roc_subclass}
therefore breaks down the ROC performance by contaminant population. The four
panels correspond, from left to right, to HSC variables, SNe in lens
galaxies, SNe in LRGs, and SNe in spiral galaxies. In
each panel, the different colors show the classifier performance at
$N_{\rm obs}=1$, 3, 7, and 14 observations. The solid curves indicate the best-seed
A2 model, while the dashed curves show the convolutional-recurrent baseline;
the shaded regions around the A2 curves represent the mean $\pm$ standard
deviation across the independent A2 runs. The A2 model already outperforms the baseline at the earliest observations, with this advantage becoming substantial at low $N_{\rm obs}$ (i.e., after only a few observations) across all four contaminant populations, which is particularly important for early detection. Even for the two most challenging contaminants, SNe in lens galaxies and SNe in LRGs, the TPR reaches $\gtrsim 70\%$ and $\gtrsim 80\%$, respectively, at $N_{\rm obs}=7$ and a target $\mathrm{FPR}\sim10^{-4}$.

The same result can also be viewed at a fixed detection efficiency.
Fig.~\ref{fig:class_fpr} shows the FPR at a fixed TPR of $80\%$, separately
for the four negative populations. The A2 model provides substantially better
false-positive control than the recurrent baseline at the same TPR, with the
FPR decreasing as observations accumulate across all four populations. The
HSC-variable and spiral-galaxy populations become much easier to reject after
only a few observations, while SNe occurring in lens galaxies and SNe in LRGs
remain the dominant contributors to the false-positive budget.

Together, these results demonstrate the practical consequence of the architectural comparison: on this LSNe dataset, the factorized attention model achieves better control of the FPR at a fixed detection efficiency, particularly during the early stages of the observation stream. The improvement is therefore not limited to a lower overall loss, but also translates into better early-detection performance.

\section{Discussion}
\label{sec:discussion}

The main lesson of this study is that the organization of attention for information exchange in the \texttt{OASIS} problem can affect performance. On the LSNe demonstration dataset, separating spatial and
temporal attention provides a more effective representation than allowing all
space--time regions to interact jointly. This difference persists across
multiple independent training runs and across the architectural variations
considered here. Restricting the spatial extent of joint attention improves its
performance substantially, suggesting that imposing some structure on how
spatial and temporal information interact is useful for this type of data.
The fact that the windowed joint models approach, but do not reach, the
factorized models is also suggestive: the useful spatial locality of the
problem can be recovered within a joint architecture, but does not by itself
reproduce the separation of spatial and temporal processing.

The readout experiments provide a second, related lesson. On the factorized
backbone, allowing the network to selectively gather information from the
causally available observations improves over treating all current spatial
features equally. A more elaborate collection of learned queries does not
provide a clear additional benefit. On the joint backbone, however, the same
selective readout provides little or no improvement. Thus, the readout cannot
be considered independently of the representation it receives: selective
weighting is useful when the preceding attention layers have organized the
information in a suitable way.

The experiments with explicit cross-band fusion lead to a similarly simple
conclusion. We do not find evidence that asynchronous multi-band data require
a dedicated color-mixing mechanism. The factorized architectures can learn
the relevant cross-band relationships through their ordinary spatial and
temporal attention. Interestingly, supplying the physical wavelengths of the
bands helps the joint architectures but has little effect on the factorized
ones. This suggests that the latter already recover much of the useful
cross-band structure from the observations themselves, although this result
should not be assumed to hold for a different set of bands or a different
science problem.

An important practical aspect of the approach is that the image encoder is
modular (Appendix~\ref{app:arch:tokens}, Fig.~\ref{fig:tokenize}). We use a
small convolutional stem trained from scratch rather than a large
pre-trained image encoder, and the demonstration does not require external
image pre-training. Whether pre-trained representations would be beneficial
for other \texttt{OASIS} applications remains an open question. Their usefulness will depend on the nature and size of the available training dataset, the images and their spatial scale, and the normalization and physical information that need to be preserved. Similarly, the appropriate cutout size and image representation are properties of the science problem rather than fixed ingredients of \texttt{OASIS}.

The losses reported here are not necessarily the lowest achievable on this dataset. In exploratory runs, a deeper convolutional stem reduced the loss of B2d from $0.1072 \pm 0.0045$ to $0.0999\pm0.0046$. In another experiment, additional normalization and augmentation reduced the loss of a small factorized model to $0.0945$. Such variations are deliberately excluded from our controlled comparison, whose purpose is to isolate the effect of the attention mechanism rather than to fine-tune each architecture for its best possible performance on this specific dataset. These experiments nevertheless indicate room for further optimization, particularly for the joint architectures, although their improved versions still do not reach the best factorized results. Finally, the data requirements of transformers may also be relevant: the available training-set size could limit performance, and larger training samples may yield further gains.

Our conclusions are necessarily limited by the demonstration. The numerical
performance and the relative ordering of architectures were obtained for the
LSNe dataset of \citet{bag2026b}; they should not be expected to transfer
unchanged to a different survey, cadence, band set, or science goal. More
broadly, important aspects of real survey streams, including long gaps,
bursts of observations, missing bands, and changes in observing conditions,
provide useful tests of the architectures presented here. These are natural
directions for future work rather than requirements of the \texttt{OASIS} formulation
itself.

\section{Conclusions}
\label{sec:conclusions}

We have formulated \texttt{OASIS} -- \emph{On-the-fly ASynchronous Image Stream} classification -- as a general problem for streams of spatially structured, multi-band observations arriving at irregular times. The formulation separates four aspects
that are often treated together: the spatial structure within each image, the
evolution of that structure with time, the asynchronous arrival of different
bands, and the requirement to make causal predictions as observations
accumulate.

Using the search for LSNe as a demonstration, we explored two broad
ways of organizing attention over such streams: factorized and joint
space--time attention. On this dataset, factorized attention provides the more
effective representation. The experiments further show that a selective
readout can improve a factorized representation, while explicit cross-band
fusion is not required. These findings illustrate how architectural choices
can be guided by the structure of the observations rather than by treating
the task as a generic classification problem.

We do not regard these results as a benchmark establishing a universally optimal architecture. The appropriate design will depend on the survey, the observing cadence, the available bands, and the scientific objective. Instead, we provide the \texttt{OASIS} formulation together with seven concrete architectures as reusable starting points for such problems. The architectures and their implementations are made publicly available so that they can be adapted and tested on other time-domain datasets and science cases. Although motivated by astronomical time-domain surveys, the underlying problem is not specific to astronomy: the same structure arises whenever observations from multiple channels arrive asynchronously, the relevant signal lies in the evolution of spatial features, and decisions must be made immediately as new observations arrive. Strong lensing is one demonstration of this broader problem, and the architectures are intended as starting points for such applications both within and beyond astronomy.

\section*{Code and data availability}

The implementations of all seven transformer-based architectures studied in this work and presented in Table~\ref{tab:headline} are publicly available on \href{https://github.com/deltasata/OASIS--On-the-fly_ASynchronous_Image-Stream-classification}{GitHub}\footnote{\url{https://github.com/deltasata/OASIS--On-the-fly_ASynchronous_Image-Stream-classification}}, together with downsampled versions of the training, validation, and test datasets used in this study. The main aim of releasing the implementations is to facilitate reuse and adaptation of the architectures, allowing them to be fine-tuned for other datasets and scientific applications. The complete simulation pipeline and full datasets are not publicly distributed because they depend on HSC-derived data products and large intermediate datasets that are impractical to redistribute. The released code provides a basis for further development and future improvements to the architectures.

\begin{acknowledgements}
Building on the authors' \texttt{ConvLSTM} baseline, initial transformer models, and experimental directions, Claude Code (Anthropic) was used in agentic mode as a development and experimentation assistant, with the aim of exploring the use of agentic AI for scientific development and automating routine implementation and experiment-management tasks. It was used to refine and fine-tune the architectures, propose and implement additional models, and run and record experiments. Claude Code also generated the \texttt{TikZ} code for the schematic diagrams through iterative refinement under the authors' supervision. Claude Code and ChatGPT (OpenAI) assisted with manuscript preparation and language editing. The scientific direction, interpretation, and conclusions remained the responsibility of the authors, who guided the model development, reviewed and verified the code, experiments, results, and manuscript, and take full responsibility for the content of this paper. We thank Daniel Gruen, Jed Homer, Lukas Heinrich, Tim Meinhardt and 
Laura Leal-Taix{\'e} for many useful discussions. This work is funded by the Deutsche Forschungsgemeinschaft (DFG, German Research Foundation) - SFB 1258 - 283604770. SB also acknowledges the support provided by the Alexander von Humboldt Foundation.
SHS thanks the Max Planck Society for support through the Max Planck Fellowship. RC acknowledges support from the French government under the France 2030 investment plan, as part of the Initiative d’Excellence d’Aix-Marseille Université - A*MIDEX (AMX-23-CEI-088). A.M. acknowledges support from FIUF137139-USACH.
\end{acknowledgements}

\bibliographystyle{aa}
\bibliography{references}

\appendix

\section{The transformer architectures in detail}
\label{app:arch}

This appendix provides the technical details of the architectures presented in the main text. It is intended for readers who wish to reproduce or adapt the models, and therefore describes implementation details omitted from the main text. Unlike a recurrent network, which processes a sequence one step at a time and carries information forward through a recurrent state, attention allows different parts of the sequence to interact directly. This provides greater flexibility in determining which information from earlier observations is relevant to the current prediction and how it should be combined with information from the current observation. The individual components are described in terms of their role in representing the observation stream, with standard ML terminology introduced where useful.

\subsection{What is attention?}
\label{app:arch:attention}

This subsection gives a compact description of the transformer block, with
particular emphasis on the attention mechanism used throughout this paper. It
is included for completeness and for readers less familiar with transformers;
readers already familiar with self-attention and transformer blocks can skip
it.

Attention \citep{Bahdanau14,Luong2015,Parikh16,vaswani2017} operates on a sequence of representations called \emph{tokens}. A token is a learned feature representation of some part of the input, and attention allows each token to selectively combine information from other tokens. This is particularly useful here because the observations form an asynchronous stream: information relevant to interpreting a new image may come from a different spatial region or an earlier observation, generally in a different photometric band.

Consider $N$ tokens, each represented by a vector of dimension $d_{\rm model}$,
and stack them into
\begin{equation}
X\in\mathbb{R}^{N\times d_{\rm model}}.
\end{equation}
An attention operation first forms three different projections of these tokens, called the queries, keys, and values, 
\begin{equation}
Q=XW_Q,\qquad K=XW_K,\qquad V=XW_V,
\end{equation}
where, for the moment,
\begin{equation}
W_Q,W_K,W_V\in
\mathbb{R}^{d_{\rm model}\times d_{\rm model}}.
\end{equation}
Thus,
\begin{equation}
Q,K,V\in\mathbb{R}^{N\times d_{\rm model}}.
\end{equation}
The query of each token is compared with the keys of all tokens, including its own. This gives the score matrix
\begin{equation}
S=QK^{\mathsf T}\in\mathbb{R}^{N\times N},
\qquad
S_{ij}=Q_iK_j^{\mathsf T}.
\end{equation}
After scaling and applying a row-wise softmax,
\begin{equation}
C=
\operatorname{softmax}_{\rm row}
\left(
\frac{S}{\sqrt{d_{\rm model}}}
\right),
\qquad
\sum_{j=1}^{N}C_{ij}=1,
\end{equation}
the new representation is
\begin{equation}
Y=CV,
\qquad
Y_i=\sum_{j=1}^{N}C_{ij}V_j.
\end{equation}
Thus $C_{ij}$ is the input-dependent weight with which token $i$ uses
information from token $j$. The weights are calculated from the current input
through the learned projections $W_Q$ and $W_K$; they are not themselves
trainable parameters.

In our application, a token represents a spatial region (patch) of an observed image, encoding its spatial features, patch position, arrival time, and photometric band.

In practice, transformers perform several such attention operations in
parallel, using different learned projections. These are called
\emph{attention heads}. With $H$ heads, each head operates in a lower
dimensional space,
\begin{equation} \label{eq:head}
d_h=\frac{d_{\rm model}}{H},
\end{equation}
with
\begin{equation}
W_Q^{(a)},W_K^{(a)},W_V^{(a)}
\in
\mathbb{R}^{d_{\rm model}\times d_h},
\qquad a=1,\ldots,H.
\end{equation}
For each head,
\begin{equation}\label{eq:mha_head}
Y^{(a)} = \operatorname{softmax}_{\rm row}
\left(
\frac{Q^{(a)}K^{(a)\mathsf T}}{\sqrt{d_h}}
\right)V^{(a)}
\in\mathbb{R}^{N\times d_h}.
\end{equation}
The outputs are concatenated and projected back to the original token
dimension,
\begin{equation}\label{eq:mha_concat}
Y=\operatorname{Concat}
\left(Y^{(1)},\ldots,Y^{(H)}\right)W_O
\in\mathbb{R}^{N\times d_{\rm model}},
\end{equation}
where
\begin{equation}\label{eq:wo}
W_O\in\mathbb{R}^{d_{\rm model}\times d_{\rm model}}.
\end{equation}

All attention operations in the architectures considered here use $H=4$
heads. Thus, for the standard $d_{\rm model}=256$ models, each head has
$d_h=64$; the wider B2w model has $d_{\rm model}=384$ and $d_h=96$.

Attention is not inherently causal. For on-the-fly classification, we therefore
restrict the allowed connections with a causal mask. If token $i$ belongs to
observation $k_i$ and token $j$ to observation $k_j$, then
\begin{equation}
C_{ij}=0
\qquad\text{for}\qquad
k_j>k_i.
\end{equation}
In practice, the corresponding entries of the score matrix are masked before
the softmax. Consequently, a prediction made after observation $k$ can depend
only on observations $1,\ldots,k$, irrespective of the attention architecture.

For compact notation, we denote the multi-head attention (MHA) operation
defined by Eqs.~\eqref{eq:head}--\eqref{eq:wo} as
\begin{equation}
\operatorname{MHA}(Q_{\rm in},K_{\rm in},V_{\rm in})
=
\operatorname{Concat}
\left(Y^{(1)},\ldots,Y^{(H)}\right)W_O,
\label{eq:mha}
\end{equation}
where $Q_{\rm in}$, $K_{\rm in}$, and $V_{\rm in}$ are the query, key, and
value inputs, respectively, and $Y^{(a)}$ is obtained from their
head-specific projections through
Eq.~\eqref{eq:mha_head}. In self-attention, the same representation provides
all three inputs, giving $\operatorname{MHA}(X,X,X)$, whereas in
cross-attention the query representation can differ from the key and value
representations. Cross-attention is discussed further in Appendix~\ref{sec:models:readout_mechanism}, where it is used in the different readout mechanisms.

A transformer block consists of an attention operation and a multi-layer perceptron (MLP), a feed-forward neural network applied independently to each token, together with residual connections and layer normalization (LN) \citep{Ba2016}. The MLP further transforms the representation produced by the attention operation. Schematically, for a post-normalized transformer block,
\begin{align}
X' &= \operatorname{LN}\!\left[
X+\operatorname{MHA}(X,X,X)
\right],\\
X'' &= \operatorname{LN}\!\left[
X'+\operatorname{MLP}(X')
\right].
\end{align}
We use post-normalization for all Class~A and B models except for B3 which uses
pre-normalization. Stacking multiple such transformer blocks progressively refines the token representation. The readout then converts the resulting representation
into the per-observation classification probability.

\subsection{From observations to tokens}
\label{app:arch:tokens}

\begin{figure*}
  \centering
  \includegraphics[width=\textwidth]{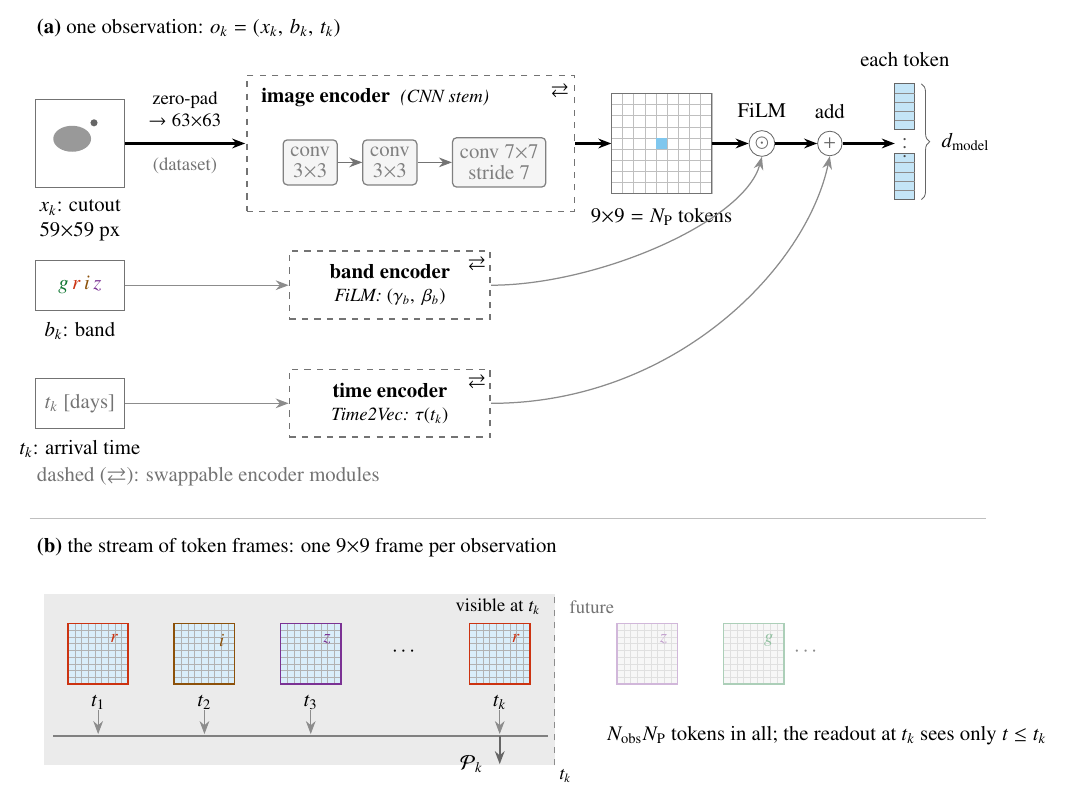}
  \caption{{\bf Top panel:} Tokenization of one observation $o_k=(x_k,\,b_k,\,t_k)$. Each part
    of the observation has its own encoder, and the three encoders (dashed)
    are modular: the image, band, or time encoder can be replaced without
    changing the attention architecture that follows. In this work, the
    image encoder is a small convolutional stem shared across bands. Its final
    layer maps each $7\times7$ spatial tile to one token
    \citep{dosovitskiy2021}, a feature vector of length $d_{\rm model}$, the
    network width. A learned vector encoding the patch position is then added
    to each token. The band encoder rescales and offsets the tokens according
    to the observed band (FiLM; \citealt{perez2018}), and the time encoder
    adds a learned representation of the arrival time (Time2Vec;
    \citealt{kazemi2019}). These operations are applied in this order, matching
    the implementation. The $59\times59$ cutout is zero-padded to
    $63\times63$ in the dataset before entering the image encoder; this is
    spatial padding and is unrelated to the missing-band padding required by
    the recurrent baseline. {\bf Bottom panel:} After tokenization, each observation is represented by a
$9\times9$ frame of tokens, with each token encoding information from its
local spatial region together with its spatial position, observation time, and
photometric band. The observations therefore form a time-ordered stream of
token frames.}
  \label{fig:tokenize}
\end{figure*}

The attention operations described above act on tokens rather than directly on
pixels. Each observation
$o_k=(x_k,\,b_k,\,t_k)$ is first converted into a set of tokens, while retaining information about where, when, and in which
band the observation was obtained. This provides the attention layers with a
representation in which spatial structure, observation time, and band
identity are all available without requiring the bands to be observed
simultaneously.

Every architecture uses the same front end, shown in Fig.~\ref{fig:tokenize}.
The image encoder is a small convolutional \emph{stem} that extracts local
image features \citep{xiao2021early}. Following the patch-based formulation
of the Vision Transformer \citep{dosovitskiy2021}, its final layer converts
the image into a spatial grid of tokens. The $59\times59$ cutout is first
zero-padded to $63\times63$ pixels and passed through two $3\times3$
convolutional layers followed by a $7\times7$ convolution with stride 7.
The resulting feature map has a $9\times9$ spatial grid, giving
$N_{\rm P}=9\times9=81$ tokens per observed image cutout. Each token is a feature
vector of length $d_{\rm model}$, which we refer to as the network
`width'. The image encoder is shared across all photometric bands.

A learned positional vector is added to each token to retain its location
within the cutout. The band information is then incorporated using feature-wise linear modulation (FiLM; \citealt{perez2018}), which applies a
band-dependent rescaling and offset to every token. Finally, the time stamp (i.e., the observation
time) is encoded with Time2Vec \citep{kazemi2019} and added to the tokens.
Thus, the same spatial representation can carry different information
depending on the band and arrival time of the observation.

For a stream containing $N_{\rm obs}$ observations, the resulting representation contains $N_{\rm obs}N_{\rm P}$ space--time tokens. In the demonstration dataset, each stream contains $N_{\rm obs}=14$ observations, and hence contains $14\times81=1134$ tokens. Note that $N_{\rm obs}$
is simply the number of observations available in the stream and is not a
fixed requirement of the architecture.

The three encoders in Fig.~\ref{fig:tokenize} are deliberately modular. The
image encoder, band encoder, and time encoder used here are one particular
implementation and can in principle be replaced independently. The attention
architectures described below operate on the resulting tokens rather than
depending on the details of these encoders.

The architectures studied below differ primarily in how attention connects the tokens. Class~A uses \emph{factorized} attention, separating spatial and temporal interactions, whereas Class~B uses \emph{joint} attention, allowing both types of interaction within the same attention operation.

\subsection{Class A: factorized (axial) attention}
\label{sec:models:A}

\begin{figure*}
  \centering
  \includegraphics[width=\textwidth]{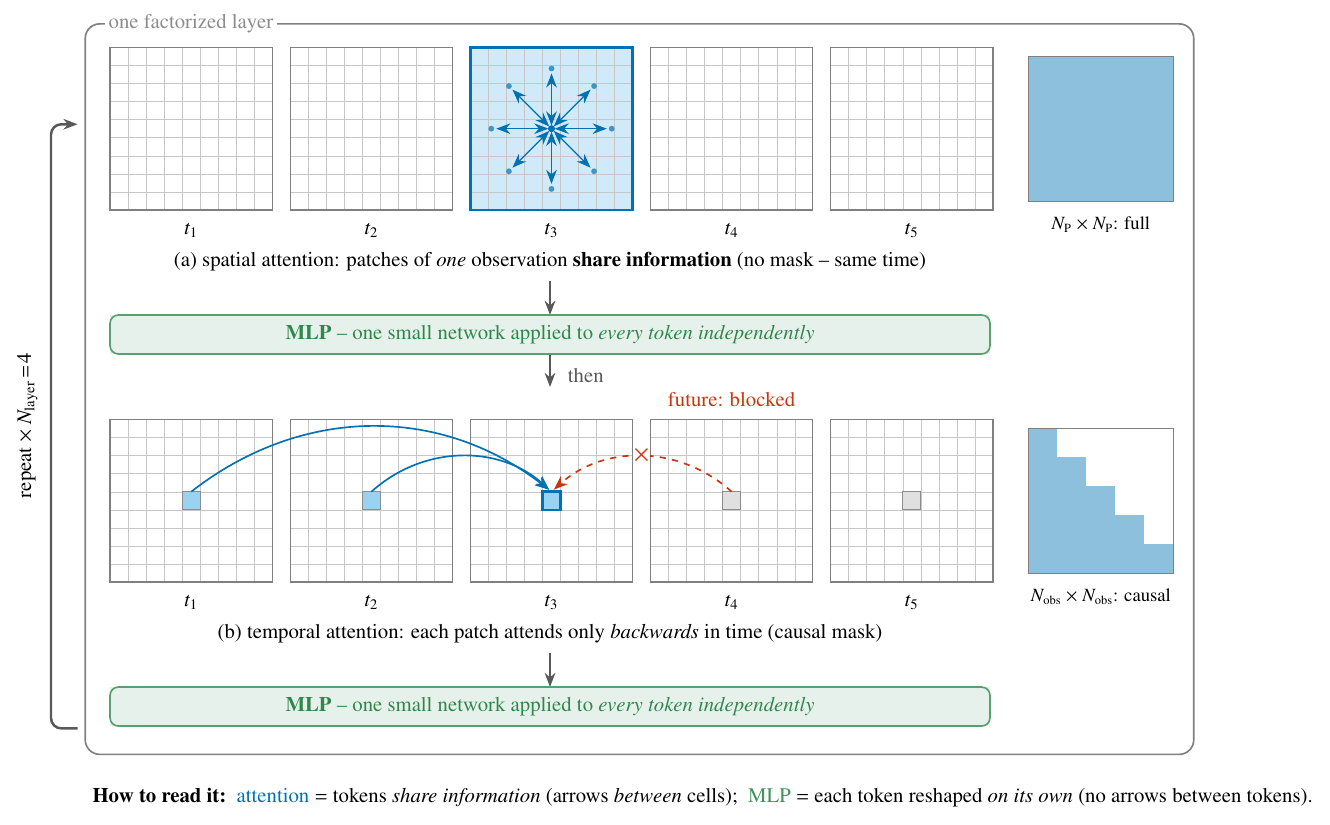}
  \caption{Class~A, factorized (axial) space--time attention. One layer applies
    (a)~full spatial attention among the $N_{\rm P}=81$ patches of a single observation
    (no mask -- same time), then (b)~strictly causal temporal attention along the
    time axis at each patch location (a token at $t$ attends only to $t'\le t$);
    each attention is followed by a per-token MLP, and the whole layer is repeated
    four times. Space and time are never mixed in a single attention.}
  \label{fig:classA}
\end{figure*}

Factorized attention \citep{ho2019,bertasius2021} separates the spatial and
temporal parts of the attention operation. In our implementation, each
factorized layer first applies full spatial attention within each observation:
the $N_{\rm P}=81$ patches of an observation can exchange information with one
another, but not with patches from other observations. This is followed by
causal temporal attention independently at each patch location, so that a
patch can attend only to the same spatial location in the current and earlier
observations. The two operations are repeated four times. All Class~A models
use $d_{\rm model}=256$ and $N_{\rm layer}=4$. Space and time are therefore
never mixed in a single attention operation, while the temporal operation is
strictly causal. Fig.~\ref{fig:classA} illustrates this arrangement.

This factorization provides a direct attention analogue of the recurrent space--time processing performed by the \texttt{ConvLSTM} baseline, while avoiding
the need to represent the asynchronous observations on a fixed grid. It also
makes the roles of the two attention operations explicit: spatial attention
combines information within an image, while temporal attention follows the
evolution of each spatial location through the observation sequence.

The Class~A architectures differ only in how the resulting space--time
representation is converted into the per-observation probability
$\mathcal{P}_k$. We refer to this operation as the \emph{readout}.

\begin{description}
  \item[A1 -- mean-pool readout.]
    The $N_{\rm P}$ tokens of the current observation are averaged with equal
    weights and passed to the shared fully connected classifier. Thus, the prediction at each
    step is based on the current observation's representation without
    selectively weighting individual tokens.

  \item[A2 -- causal cross-attention readout.]
    A single learned \emph{query} vector is used to gather information from
    all causally available space--time tokens. In attention terminology, the
    query specifies what information to extract, while the attention weights
    determine how strongly each token contributes. The query is combined with
    the time encoding for the current observation, so the readout can produce
    a different weighting as the sequence evolves. This form of attention pooling with a learned query is motivated by  \citet{lee2019set}. Unlike A1, the readout can selectively use information
    from earlier observations as well as the current one.

  \item[A3 -- latent-array readout.]
    A3 extends A2 by replacing the single query with an array of $16$ learned
    latent vectors for each observation, in the spirit of the latent
    bottleneck of the Perceiver architecture \citep{jaegle2021}. The latent vectors are refined over
    three rounds, each consisting of causal cross-attention from the latents
    to the visible space--time tokens, self-attention among the $16$ latents,
    and a feed-forward layer (MLP). The refined latents are then averaged and passed
    to the classifier. This provides a richer version of the selective
    readout, while retaining the same factorized backbone. The latent array
    therefore acts only at the readout stage; the space--time representation
    itself is still formed by the factorized attention layers and is not
    forced through this latent bottleneck.
\end{description}

\subsection{Class B: joint attention}
\label{sec:models:B}

\begin{figure*}
  \centering
  \includegraphics[width=\textwidth]{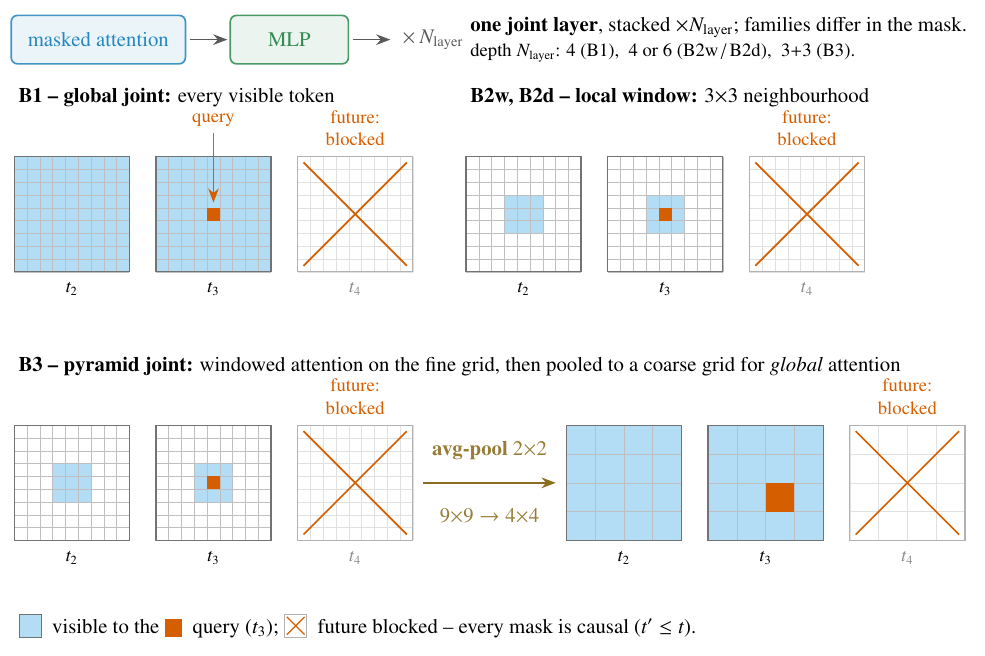}
  \caption{Class~B, joint space--time attention over all $14\times81=1134$
    tokens, in its three variants. Shaded cells mark the tokens visible to the
    query token (orange), drawn at $t_3$; the greyed-out observation $t_4$
    lies in the future and is blocked in every variant -- all masks are
    causal. Restricting the spatial receptive field is what makes joint
    attention competitive: the global variant B1 is diluted and fails to
    beat the \texttt{ConvLSTM}+\texttt{LSTM} baseline, the $3\times3$-window family (B2w wide,
    B2d deep) is the best joint solution, and the pyramid B3 -- windowed on
    the fine grid, global on the pooled coarse grid -- recovers global
    context but does not improve on the window.}
  \label{fig:classB}
\end{figure*}

Joint attention applies a single attention operation to all space--time tokens,
with a causal mask that allows a token at observation $t$ to attend to every
token at $t'\le t$. Unlike Class~A, spatial and temporal information can
therefore interact directly within the same attention operation. 
We include it as a natural alternative to the factorized organization and
vary its spatial receptive field to examine how much freedom of interaction is
useful.

\begin{description}
  \item[B1 -- global joint.]
    Every causally visible token can attend to every other visible token
    (width $256$, four layers). This is the unrestricted joint architecture
    and provides the reference case for this class.

  \item[B2w, B2d -- local-window joint.]
The spatial receptive field is restricted to a $3\times3$ patch window
\citep[in the spirit of][]{liu2021}, while attention remains causal in
time. A token at position $p$ can therefore attend to the corresponding
$3\times3$ spatial neighborhood in the current and all previous
observations, but not to more distant spatial positions or future
observations. We use two capacity choices: B2w is wider
($d_{\rm model}=384$, four layers), while B2d is deeper
($d_{\rm model}=256$, six layers). Comparing them tests whether the
behavior of the local-window architecture depends strongly on model
width or depth.

  \item[B3 -- hierarchical pyramid joint.]
Joint attention is first restricted to a local $3\times3$ window on the
fine $9\times9$ spatial grid, as in B2w and B2d, while retaining the
causal temporal mask. Thus, each token can attend to the corresponding
local spatial neighborhood across the current and all previous
observations. After three such layers, the resulting feature maps are
average-pooled over $2\times2$ spatial regions, reducing the grid from
$9\times9$ to $4\times4$. Joint attention is then applied without spatial
restriction on this coarse grid, again with the causal temporal mask,
for another three layers. Thus, B3 combines local fine-scale and global
coarse-scale space--time interactions in a hierarchical architecture.
By applying global attention only after spatial pooling, B3 provides access to larger-scale spatial context without applying global attention directly to all $9\times9\times14=1134$ fine-grid space--time tokens, as in B1.
\end{description}

Fig.~\ref{fig:classB} illustrates the spatial receptive fields and causal
masks of the three joint architectures. All joint models use mean pooling for readout, except B3, which uses single-query cross-attention with the current time embedding as the query.

\subsection{Readout}
\label{sec:models:readout}

\begin{figure*}[h]
  \centering
  \includegraphics[width=\textwidth]{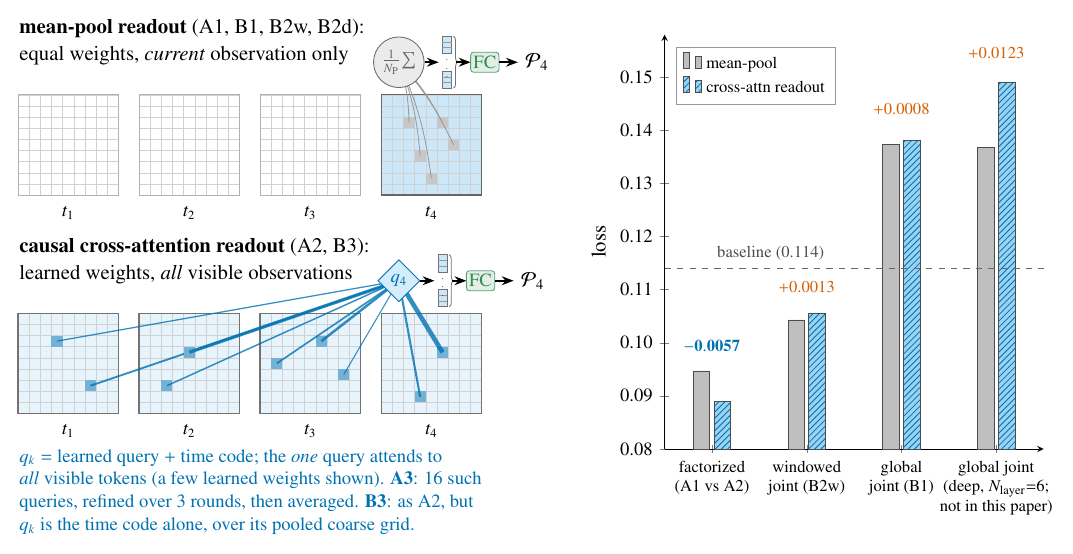}
  \caption{The two readouts, and the asymmetry of their effect. \emph{Left:}
    the mean-pool readout (used in A1, B1, B2w and B2d models) averages the $81$ tokens of the current observation with equal weights; the causal cross-attention readout (used in A2, A3 and B3 models) forms one or more queries and attends, with learned weights, over \emph{every} causally visible token. A2 uses a single query formed from a learned vector plus the time encoding, whereas A3 uses sixteen such queries per step, refined together and then averaged. B3, on the other hand, uses a single query given solely by the time encoding.
    In all cases, the pooled tokens form one $d_{\rm model}$ vector that a shared
    fully-connected head and a sigmoid map to $\mathcal{P}_k$; only the pooling
    differs. \emph{Right:} the same cross-attention readout that improves the factorized
    model (A1$\to$A2) leaves the windowed joint (B2w) and global joint (B1)
    essentially unchanged, and clearly harms a deeper global variant (six
    layers, not part of this paper); the dashed line marks the \texttt{ConvLSTM}+\texttt{LSTM}
    baseline.}
  \label{fig:readout}
\end{figure*}

After the attention backbone has exchanged information among the causally available tokens, the resulting representation retains the frame structure illustrated in the bottom panel of Fig.~\ref{fig:tokenize}, with a $9\times9$ spatial representation for each observation in all models except B3, which uses a $4\times4$ representation. Each token can now carry information from other spatial patches and earlier observations. The readout is an independent architectural choice that converts this
space--time representation into the per-observation prediction, $\mathcal{P}_k$. The same readout mechanisms can therefore be attached to
either the factorized or joint attention classes. We consider three choices:
mean pooling, a single-query causal cross-attention readout, and a
latent-array readout with sixteen learned queries. The latter two are
selective readouts: rather than assigning equal weight to the available
tokens, they learn how much information to extract from each causally visible
token.

The three readouts are briefly described in Appendix~\ref{sec:models:A};
the details are given below for readers interested in the implementation.
Fig.~\ref{fig:readout} illustrates the mean-pool and single-query readouts. The latent-array readout used by A3 extends the latter from one learned query to sixteen learned queries that are refined together. The single-query readout used by A2 combines a learned query with the time embedding of the current observation, whereas B3 uses the time embedding itself as the query. The effect of the readout therefore depends on the preceding attention backbone, as discussed in Sec.~\ref{sec:results}.

\subsubsection{Readout mechanisms}
\label{sec:models:readout_mechanism}

After the attention backbone, each spatial patch $p$ at observation $t$ is
represented by a token
$x_{t,p}\in\mathbb{R}^{d_{\rm model}}$. At observation $k$, the indices of the causally available tokens are
\begin{equation}
\Lambda_k =\{(t,p):t\leq k,\;p=1,\ldots,N_{\rm P}\}.
\label{eq:readout_cone}
\end{equation}

\paragraph{Mean pooling.}
The mean readout averages the spatial tokens of the current observation,
\begin{equation}
h_k =
\frac{1}{N_{\rm P}}\sum_{p=1}^{N_{\rm P}}x_{k,p}.
\label{eq:readout_mean}
\end{equation}
Information from earlier observations can therefore enter $h_k$ only through
the causal attention in the backbone.

\paragraph{Single-query causal cross-attention.}
The cross-attention readout uses a learned query to selectively extract
information from the causal space--time representation, a form of attention
pooling with a learned query \citep{lee2019set}. For the factorized
models,
\begin{equation}
q_k=r+\tau_k,
\qquad
\tau_k\equiv\tau(t_k),
\label{eq:readout_query}
\end{equation}
where $r\in\mathbb{R}^{d_{\rm model}}$ is learned and $\tau_k$ is the
Time2Vec embedding of the current observation time.

Unlike self-attention, in which the queries, keys, and values are derived
from the same representation, cross-attention allows them to come from different
representations. Here the readout query provides $Q$, while the causally
available backbone tokens provide $K$ and $V$. Thus, for the $H$ attention heads,
\begin{equation}
Q_k^{(a)}=q_kW_Q^{(a)},\qquad
K_{t,p}^{(a)}=x_{t,p}W_K^{(a)},\qquad
V_{t,p}^{(a)}=x_{t,p}W_V^{(a)},
\label{eq:readout_qkv}
\end{equation}
where $a=1,\ldots,H$ and
$d_h=d_{\rm model}/H$. As in the transformer blocks described above, we use $H=4$ attention heads. The attention weights over the causal cone are
\begin{align}
c_k^{(a)}(t,p)
&= \operatorname{softmax}_{(t,p)\in\Lambda_k}\left[\frac{Q_k^{(a)}{K_{t,p}^{(a)}}^{\mathsf T}}{\sqrt{d_h}}\right] \nonumber\\
&= \frac{\exp\!\left(Q_k^{(a)}{K_{t,p}^{(a)}}^{\mathsf T}/\sqrt{d_h}\right)}
{\displaystyle\sum_{(t',p')\in\Lambda_k}\exp\!\left(Q_k^{(a)}{K_{t',p'}^{(a)}}^{\mathsf T}/\sqrt{d_h}\right)}\;.
\label{eq:readout_weights}
\end{align}
The causal restriction is imposed by excluding tokens with $t>k$ before the
softmax. 
For the single-query readout, the output of attention head $a$ is the
$d_h$-dimensional vector
\begin{equation}
y_k^{(a)} =
\sum_{(t,p)\in\Lambda_k}
c_k^{(a)}(t,p)V_{t,p}^{(a)} \in\mathbb{R}^{d_h}.
\label{eq:readout_head}
\end{equation}
This is the single-query analogue of $Y^{(a)}$ in Eq.~\eqref{eq:mha_head}.
The outputs of the $H$ heads are then concatenated and projected to form the readout vector,
\begin{equation}
h_k=
\operatorname{Concat}
\left(y_k^{(1)},\ldots,y_k^{(H)}\right)W_O\in\mathbb{R}^{d_{\rm model}}\;,
\label{eq:readout_hk}
\end{equation}
where
\begin{equation}
W_O\in\mathbb{R}^{d_{\rm model}\times d_{\rm model}}
\end{equation}
is the output projection matrix. There is no residual connection or MLP in this readout. The resulting vector is layer-normalized and passed to the common
classifier.

Following the notation of Eq.~\eqref{eq:mha}, the multi-head cross-attention readout can be written compactly as
\begin{equation}
h_k=\operatorname{MHA}\!\left(q_k,X_{\Lambda_k},X_{\Lambda_k}\right),
\label{eq:readout_mha}
\end{equation}
where the single query $q_k$ provides the query input, while
$X_{\Lambda_k}\equiv\{x_{t,p}:(t,p)\in\Lambda_k\}$ denotes the causally
available backbone tokens used as the key and value inputs.

The same cross-attention readout is used by B3, except that
$q_k=\tau_k$ and the keys and values are taken from its coarse $4\times4$ spatial representation across the observation sequence.

\paragraph{Latent-array readout.}
The latent-array readout replaces the single query with $16$ learned
latent vectors, inspired by the latent bottleneck introduced
in the Perceiver architecture \citep{jaegle2021}. The learned vectors are shared across observations, while
the current time embedding makes their initial state vary across observations:
The initial latent array for observation $k$ is
\begin{equation}
\ell_k^{(0)}=
\begin{bmatrix}
\ell_{k,1}^{(0)}\\
\vdots\\
\ell_{k,16}^{(0)}
\end{bmatrix}
\in\mathbb{R}^{16\times d_{\rm model}},
~~~
\ell_{k,j}^{(0)}=\ell_j+\tau_k,
~~~ j=1,\ldots,16,
\label{eq:readout_latent_initial}
\end{equation}
where $\ell_j$ are learned latent vectors shared across observations and
$\tau_k$ is the Time2Vec embedding of the current observation time. The latent
array is then refined through three successive blocks, each with independent
parameters. Within the first block,
\begin{align}
\ell_k^{(0)\prime} &= \ell_k^{(0)}+\operatorname{MHA}\!\left[\operatorname{LN}(\ell_k^{(0)}),X_{\Lambda_k},X_{\Lambda_k}\right], \label{eq:l0p}\\
\ell_k^{(0)\prime\prime} &= \ell_k^{(0)\prime}+\operatorname{MHA}\!\left[\operatorname{LN}(\ell_k^{(0)\prime}),\operatorname{LN}(\ell_k^{(0)\prime}),\operatorname{LN}(\ell_k^{(0)\prime})\right],\label{eq:l0pp}\\
\ell_k^{(1)} &= \ell_k^{(0)\prime\prime}+\operatorname{MLP}\!\left[\operatorname{LN}(\ell_k^{(0)\prime\prime})\right]\label{eq:l1}\;,
\end{align}
where $\operatorname{LN}$ denotes layer normalization and $\operatorname{MHA}$
follows the compact notation for multi-head attention introduced in
Eq.~\eqref{eq:mha}. The first operation,
Eq.~\eqref{eq:l0p}, is causal cross-attention from the latent array to the
backbone tokens in $\Lambda_k$. The second, Eq.~\eqref{eq:l0pp}, is
self-attention among the 16 latents at the same observation, while the final
operation, Eq.~\eqref{eq:l1}, applies an MLP independently to each latent.
The output $\ell_k^{(1)}$ of the first block is then used as the input to the
second block, and the output of the second block is similarly passed to the
third. The attention and MLP layers in the three blocks have separate weights,
while the same learned latent vectors $\ell_j$ are used to initialize the
array at every observation. After the third block, the 16 refined latents in
$\ell_k^{(3)}$ are averaged,
\begin{equation}
h_k=\frac{1}{16}\sum_{j=1}^{16}\ell_{k,j}^{(3)}
\in\mathbb{R}^{d_{\rm model}},
\label{eq:readout_latent_mean}
\end{equation}
and the resulting $h_k$ is passed to the common classifier.

For all readout mechanisms, the resulting vector $h_k\in\mathbb{R}^{d_{\rm model}}$
is passed to a shared fully connected classifier, as illustrated by the ``FC'' block in Fig.~\ref{fig:readout}, consisting of two linear
layers with a GELU activation and dropout between them. The resulting scalar
logit is then mapped to the classification probability $\mathcal{P}_k$ with a sigmoid.

\subsection{Training protocol}
\label{app:arch:training}

All runs minimize the per-step binary cross-entropy loss using the \texttt{Adam} optimizer \citep{kingma2015}, with a batch size of $128$ samples and a dropout rate of $0.1$. The band and
time encoders are those described in Appendix~\ref{app:arch:tokens}. The
learning rate follows a cosine schedule with periodic resets to its maximum
value (``warm restarts''; \citealt{loshchilov2017}). Training uses a patience
rule: after 32 passes over the training set without improvement in the validation loss, the run is
rolled back to its best state and training is continued.

To reduce the dependence on any particular random initialization, each
architecture in Table~\ref{tab:headline} was trained from more than ten
independent random seeds, changing only the seed. The mean and standard
deviation reported in the table summarize these runs.

\end{document}